\documentclass{article}

\usepackage{arxiv}

\usepackage[utf8]{inputenc} 
\usepackage[T1]{fontenc}    
\usepackage{hyperref}       
\usepackage{url}            
\usepackage{booktabs}       
\usepackage{amsfonts}       
\usepackage{nicefrac}       
\usepackage{microtype}      
\usepackage{lipsum}
\usepackage{graphicx}
\graphicspath{ {./images/} }
\usepackage{siunitx}

\usepackage{xcolor}

\title{Experimental measurements of the transfer function of a nonlinear optical loop mirror}

\author{
Alix Malfondet \\
Laboratoire Interdisciplinaire Carnot de Bourgogne\\
UMR 6303 CNRS Université Bourgogne Franche-Comté\\
9 Av. A. Savary, B.P. 47869, 21078 Dijon Cedex, France \\
   \And
Alexandre Parriaux \\
Laboratoire Interdisciplinaire Carnot de Bourgogne\\
UMR 6303 CNRS, Université Bourgogne Franche-Comté\\
9 Av. A. Savary, B.P. 47869, 21078 Dijon Cedex, France \\
  \And
Patrice Tchofo-Dinda \\
Laboratoire Interdisciplinaire Carnot de Bourgogne\\
UMR 6303 CNRS, Université Bourgogne Franche-Comté\\
9 Av. A. Savary, B.P. 47869, 21078 Dijon Cedex, France \\
  \And
Guy Millot \\
Laboratoire Interdisciplinaire Carnot de Bourgogne\\
UMR 6303 CNRS, Université Bourgogne Franche-Comté\\
9 Av. A. Savary, B.P. 47869, 21078 Dijon Cedex, France \\
Institut Universitaire de France (IUF), 1 Rue Descartes, Paris, France\\
}

\begin{document}
\maketitle

\begin{abstract}
Measurement of the average values of the input and output powers of a device can give insight into the transfer function (TF) of that device, but this approach usually hides the real impact of certain propagation phenomena. 
However, to the best of our knowledge, measurements of the TF of nonlinear optical loop mirrors have always been carried out using this approach~\cite{Stephan2011,Rafidi2012,Wen2017}, which may lead to underestimating the impact of dispersive and nonlinear effects.
Here we present the experimental measurement of the TF of a nonlinear optical loop mirror (NOLM), made from the experimental measurements of the input and output intensity profiles of the device, using a frequency-resolved optical gating (FROG) technique~\cite{Kane1993,DeLong1994,Dudley2001}. Our approach clearly highlights the impact of dispersion effects and Kerr nonlinearity on the NOLM's TF.
\end{abstract}

\section{Introduction}
\label{Introduction}

\textcolor{black}{
The NOLM is an optical device highly appreciated for its versatility and for the diversity of its applications. Some optical systems use the NOLM in its passive version (without active component) \cite{Doran1988,Blow1989,Blow1990,Labruyere2006,Ikeda2006,Li2014,Malfondet2021} while its active version (including an optical amplifier) is better suited to other systems \cite{Fermann1990,Duling1994}.  }
Through a proper adjustment of its TF, the NOLM can be designed for a variety of applications, such as switching, mode-locking, pulse regeneration, analog to digital converter, noise suppression, or multiplexing~\cite{Blow1989,Blow1990,Labruyere2006,Duling1994,Ikeda2006,Fermann1990,Li2014,Perego2019,Guo2019,Wen2019}. 
To perform all these operations, the TF of the NOLM, defined as the functional dependence of the device's output power upon the input power, is essential.
For instance, the TF must have a saturation effect for NOLM to be used as a saturable absorber. However, to be used for pulse regeneration, the slope of the TF at the origin must be as low as possible, which is necessary to suppress the noise~\cite{Ludwiga}. 
The NOLM is therefore a versatile device, which must be properly configured to be used effectively in the desired application. In this context, an accurate measurement of its instantaneous TF is essential.

\textcolor{black}{
One of the main components of the NOLM is its all-fiber loop. The length of the loop can vary greatly depending on the application the NOLM is designed for, and typically ranges from a few meters to a few kilometers. Thus, in the vast majority of applications, the length of the loop is such that the effects of dispersion and non-linearity inevitably have an impact on the temporal profile of the light beams passing through the NOLM. However, it is well known that the impact of dispersion effects strongly depends on the temporal profile of the light field, as it is generally negligible for nanosecond pulses but very detrimental for ultra-short pulses of the order of a few picoseconds. The combined effects of dispersion and non-linearity on ultra-short pulses are difficult to detect with conventional techniques for measuring the transfer function, i.e., without devices for characterizing the pulse temporal profile. In the present work, we unveil an experimental technique for measuring the instantaneous transfer function of a NOLM operating with ultra-short light pulses. }

In our experiments, the TF is obtained by measuring the temporal profiles of the input and output pulses of the NOLM, using a multi-shot FROG system provided by the company Femto-Easy. This apparatus is comparable to an intensity autocorrelator, whose photo-detector is replaced by a spectrograph. A nonlinear crystal, designed to perform second-harmonic generation (SHG), is also placed before the detector.
The signal obtained by the SHG process is then measured, and its intensity is recorded as a function of the time delay and optical frequency.
This measurement, called 'spectrogram' (or 'FROG trace'), is used as input data by an algorithm to retrieve the pulse temporal profiles that we use to plot the TF of the NOLM~\cite{Trebino1997,Trebino2000}.
Thus, here the FROG technique does not provide a direct measurement of the intensity profile, unlike what can be obtained with the combination of a photodiode and an oscilloscope. The choice of the FROG system is justified by the short duration of our pulses (0.75 to \SI{3.4}{\pico\second}), which are too short to be properly measured with a photodiode.

An effect capable of significantly distorting the TF of the NOLM is the chromatic dispersion of the fiber constituting the NOLM's loop, and more specifically, the effect of the third-order dispersion (TOD). Indeed, it is well known that a fiber with non-zero TOD generates an asymmetry in the temporal profile of the pulse.
In principle, this asymmetric distortion of the pulse profile is more pronounced when the fiber length is large or the TOD coefficient is large.
Asymmetric distortions in pulse profiles are impossible to measure with an autocorrelator, while our FROG-based approach is suitable for any type of pulse profile~\cite{Trebino2000}.

\section{Measurement of the transfer function}

\subsection{NOLM architecture}
\label{sec:NOLM}

\begin{figure}[h]
\centering\includegraphics[width=.5\linewidth]{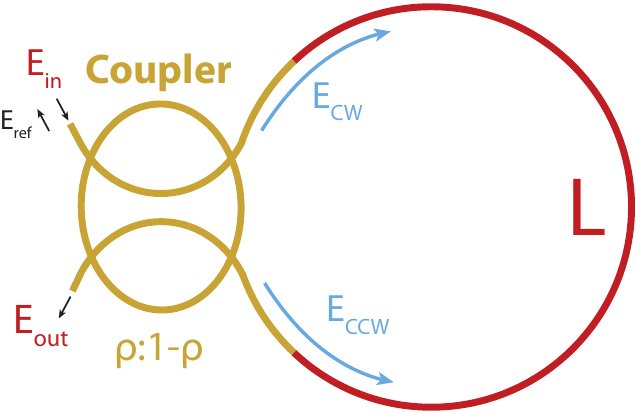}
\caption{Schematic of a conventional NOLM}
\label{fig:nolm}
\end{figure}

The architecture of a conventional NOLM corresponds to a relatively simple device, which consists in connecting the two output ports of a bidirectional coupler to the two ends of a fiber section, so as to form a loop, as illustrated schematically in Fig. \ref{fig:nolm}.
The characteristics of the section of fiber constituting the NOLM loop, namely its length, Kerr nonlinearity coefficient, second- and third-order dispersion coefficients, and birefringence, are the main parameters structuring the TF of a NOLM. However, fiber length and Kerr nonlinearity are the only characteristics that come into play to structure the TF of an ideal NOLM.
The other mentioned characteristics generate disruptive phenomena causing undesirable distortions in the TF. In other words, the NOLM's loop fiber must be chosen to have characteristics that minimize as much as possible the effects of dispersion and birefringence, while having the highest possible nonlinearity to keep its length as short as possible.

The fiber that seems to us to achieve the best compromise between the constraints raised is a commercial fiber with a nonlinearity coefficient of \SI{20}{\watt^{-1}\kilo\meter^{-1}} and relatively low second- and third-order dispersion coefficients, equal to \SI{0.77}{\pico\second^{2}\kilo\meter^{-1}} and \SI{4.1e-2}{\pico\second^{3}\kilo\meter^{-1}}, respectively. From now on, this commercial fiber is called LD-NLF (Low-Dispersion Highly Nonlinear Fiber).

The architecture of our NOLM is based on that of a conventional NOLM, but where we use an adjustable ratio coupler instead of a standard fixed ratio coupler. The use of a tunable coupler allows flexibility in setting the TF of the NOLM.
\color{black}
It is easy to get a rough idea of the TF of a NOLM in the case where the incident light is a continuous wave or a pulse of relatively long duration, i.e. in situations where the chromatic dispersion effects are negligible. To this end, it should be noted that the role of the coupler is to separate the incident beam into two beams ($E_{cw}$ and $E_{ccw}$) and inject them respectively at both ends of the loop-shaped fiber of the NOLM, as schematically illustrated in Fig.\ref{fig:nolm}. The two beams then propagate in the opposite direction in the looped fiber. 
Initially, the electric fields for the clockwise and counter-clockwise beams, $E_{cw}$ and 
$E_{ccw}$, are given by $E_{{cw}_0}=\sqrt{\rho}E_{in}$  and $ E_{{ccw}_0}=i\sqrt{1-\rho }E_{in} $
where $E_{in}$ is the electric field of the incident beam and $\rho$ denotes the coupler ratio, which is an adjustable parameter in our NOLM.
Taking into account only the self-phase modulation effect, after propagation in the NOLM loop (of length L), the expressions of the field amplitudes become:

\begin{equation}    
E_{cw}  = E_{{cw}_0} \exp(i\gamma L |E_{{cw}_0}|^2 ) = \sqrt{\rho}E_{in} \exp(i\gamma L \rho |E_{in}|^2 ) \label{Ecw}
\end{equation}
\begin{equation}    
E_{ccw} = E_{{ccw}_0} \exp(i\gamma L |E_{{ccw}_0}|^2 )= i\sqrt{1-\rho }E_{in} \exp[i\gamma L (1-\rho) |E_{in}|^2 ]  \label{Eccw} 
\end{equation}

Then, each of the two beams exits the loop by passing again through the coupler, which breaks it into two fractions distributed respectively over the input and output ports.
On each of the two ports, two fractions of beams originating respectively from the fields $ E_ {cw} $ and $ E_ {ccw} $ combine coherently, so as to give respectively a reflected beam and an output beam whose fields are given by:

\begin{eqnarray}
E_{ref}  =\sqrt{\rho}E_{ccw} +  i\sqrt{1-\rho }E_{cw},
\label{Eref}\\
E_{out} = \sqrt{\rho}E_{cw}+ i\sqrt{1-\rho }E_{ccw}.
\label{Eout}
\end{eqnarray}
The corresponding powers are given by:
\begin{eqnarray}
P_{ref}  \equiv |E_{ref}|^2 =   A (1+\text{cos}(B P_{in})P_{in},
\label{Pref}\\
	P_{out}  \equiv |E_{out}|^2 = [1-A (1+\text{cos}(B P_{in})]P_{in},
\label{Pout}
\end{eqnarray}
with $A\equiv 2\rho(1-\rho)$ and $B\equiv \gamma L (1-2\rho)$.

If necessary (i.e., in case of relatively long loop fiber or high attenuation coefficient), the attenuation effect can be easily taken into account, by using the following expression:
\begin{eqnarray}
P_{out}  = \Gamma  [1-A (1+\text{cos}(B_{eff} P_{in})]P_{in},   \label{eq_tf}
\end{eqnarray}
	with $B_{eff}\equiv \gamma L_{e}(1-2\rho)$,  $L_{e}\equiv  (1-\Gamma)/\alpha$ and $\Gamma \equiv \text{exp}(-\alpha L) $,  where  $L_{e}$ its effective length of the looped fiber and $\alpha$ the attenuation coefficient including linear and connection losses.

From now on, we will refer to the procedure presented above (to obtain the analytical expression of the output power of the NOLM as a function of the input power) as being the conventional procedure for determining the TF of the NOLM.
In our NOLM, the reflected power $P_{ref}$ is suppressed by means of a circulator
placed just before the entry port of the NOLM. Moreover, the expression (\ref{Pout}) illustrates that the output power of the NOLM does not vary monotonically with the input power, but rather in oscillatory way.
\color{black}
 In other words, the TF of a NOLM is basically an interference pattern, where two waves combine coherently with a phase shift created by the self-phase modulation effect.
Most practical applications of NOLMs only use the first bright fringe of this interference pattern. Taking into account the non-linearity coefficient of the loop fiber indicated above, we have chosen a fiber length of 20m, in order that the self-phase modulation effect is sufficient for the NOLM to operate on the first bright fringe of its interference pattern, with a saturation effect.

\subsection{Experimental setup}
\label{sec:exp_setup}

\begin{figure}[ht]
\centering\includegraphics[width=.85\linewidth]{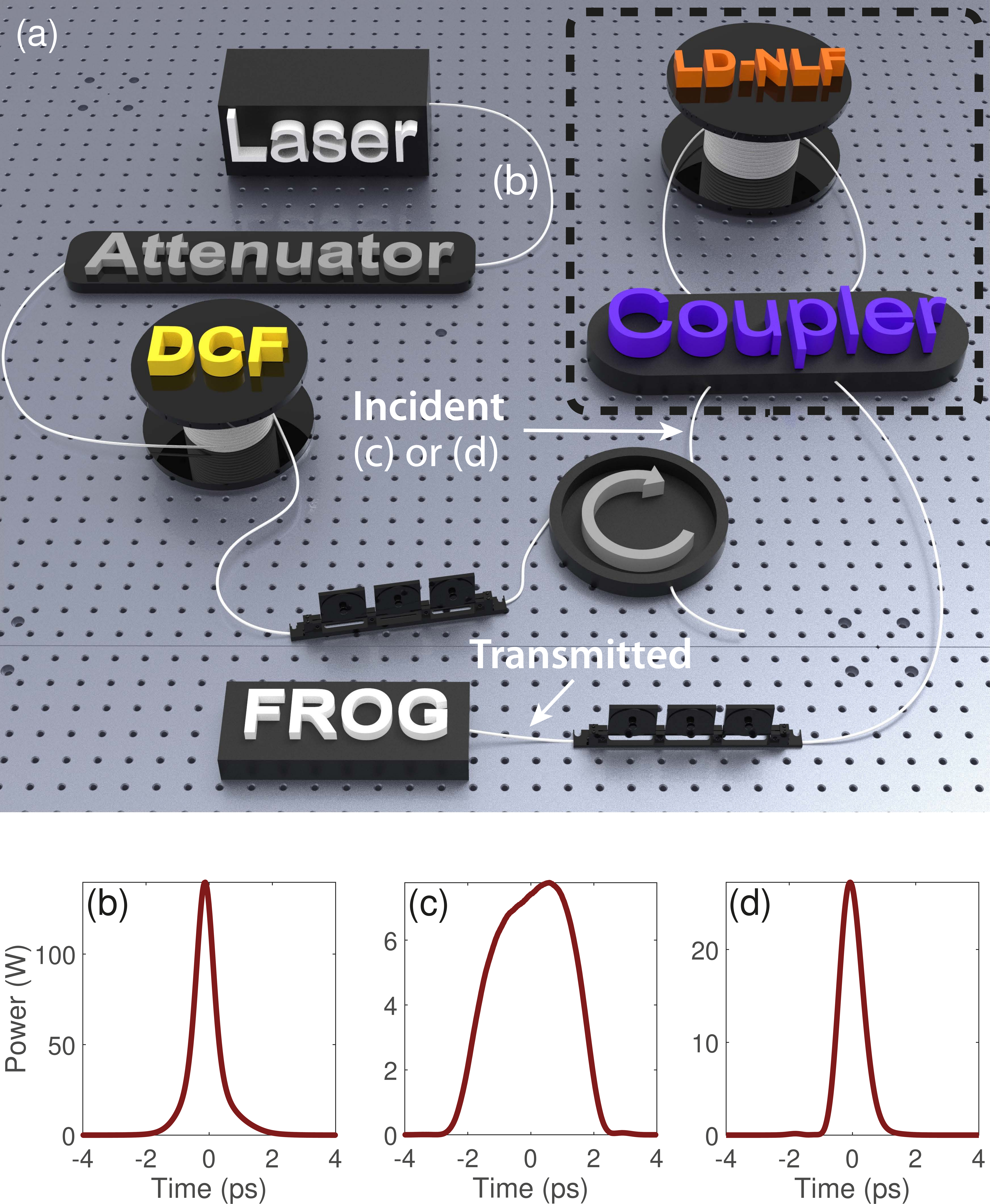}
\caption{(a) Experimental setup used for the measurement of the TF of the NOLM. DCF : Dispersion compensation fiber, LD-NLF : Low-dispersion nonlinear fiber, BPF : Band-pass filter, FROG : Frequency-resolved optical gating. (b) Signal at the output of the laser source. (c) Incident signal broadened by the DCF. (d) Incident signal if we remove the DCF.}
\label{fig:setup}
\end{figure}

In most practical applications using NOLMs, the intensity profile of the light passing through the NOLM may include ultra-fast modulations at frequencies that can reach the Terahertz, or ultrashort pulses with temporal widths as short as one picosecond. 
In our TF measurement experiments, the light entering the NOLM is a pulse with a
 temporal width of the order of a picosecond.
Our experimental setup for measuring the NOLM TF is depicted in Fig. \ref{fig:setup} \hyperref[fig:setup]{(a)}, in which the NOLM is delimited by the dotted frame. The places of measurement of the input and output signals of the NOLM are clearly indicated in this figure. The pulses injected into the NOLM are generated by a laser source that delivers \SI{0.75}{\pico\second} pulses at a repetition frequency of \SI{40}{\mega\hertz}, with an average power of \SI{5}{\milli\watt}. The temporal profile of these pulses, retrieved with the FROG technique, is shown in Fig. \ref{fig:setup} \hyperref[fig:setup]{(b)}. A \SI{7}{} meter dispersion compensating fiber (DCF) is used to broaden the pulses in order to limit asymmetric distortions induced by the TOD. As we can see in Fig. \ref{fig:setup} \hyperref[fig:setup]{(c)}, the input signal is slightly distorted after passing through the DCF, due to the combined effects of dispersion and non-linearity~\cite{Agrawal2019}. A variable attenuator is placed at the output of the laser to reduce the peak power of the pulses, and to limit nonlinear effects occurring inside the DCF. A circulator is used to protect the laser from the signal reflected by the NOLM.

In our experimental setup we do not use polarization maintaining (PM) components, 
but we use a polarization controller (PC) at the input of the circulator in order 
to reduce the impact of birefringence effects.
 We indeed adjust this PC to be as close as possible to the scalar configuration considered in our simulations.  Another PC is used at the input of the FROG system
 in order to optimize the SHG signal. However, it should be noted that previous studies have shown that the NOLM can also fully operate with PM elements, for applications such as mode-locking or time-division demultiplexing~\cite{Sakamoto2002,Szczepanek2015,Pielach2020}.

\subsection{Methodology}

We now present our methodology for measuring the NOLM TF, and our numerical modeling procedure. First, the average power of the incident signal is adjusted using the variable attenuator to be equal to \SI{1}{\milli\watt}. 
The peak power of the incident signal is then equal to \SI{28}{\watt} when the DCF is 
not included in the cavity, and \SI{7.8}{\watt} when the DCF is inserted into
 the cavity.
	The temporal measurement of the input and output signals are performed using a multi-shot FROG system operating with a signal recovery algorithm. The latter yields an average measurement over a few hundred or thousand of pulses that have been acquired during the scan. The average power of the signal is also necessary to reconstruct the pulse temporal profile, because the FROG system can only give a normalized intensity profile. 

The main system parameters having the most impact on the NOLM TF, as well as the physical phenomena involved in the structuring of the TF, can be identified using a light propagation model for the NOLM.
The system consists of lumped components (couplers, connectors, splices), and passive fiber sections comprising the loop fiber of the NOLM and a few small pieces of standard SMF fibers that serve as cords in the component connectors. The lumped elements only affect the pulse amplitude linearly. The scalar pulse propagation in the passive fibers can be described by means of the following scalar  nonlinear Schrödinger equation:

\begin{equation}
\label{eqn:NLSE}
\mathrm{i} \frac{\partial \psi}{\partial z}=\frac{\beta_{2}}{2} \frac{\partial^{2} \psi}{\partial t^{2}}+\frac{\mathrm{i} \beta_{3}}{6} \frac{\partial^{3} \psi}{\partial t^{3}} - \gamma|\psi|^{2}\psi -\frac{\alpha}{2} \psi
\end{equation}
where $\beta_{2}$ is the second order dispersion coefficient, $\beta_{3}$ the TOD coefficient, $\gamma$ the nonlinear coefficient, and $\alpha$ the attenuation coefficient. 

\textcolor{black}{
Here, to model the action of the NOLM, we do not follow the conventional procedure presented in the section \ref{sec:NOLM}, because that procedure is only suitable for situations where the incident light corresponds to a continuous wave, or to pulses of relatively long duration. In the present work, we consider rather situations where the incident light is an ultra-short pulse with a duration of the order of a few picoseconds. Knowing that such pulses are very sensitive to dispersion effects, it can be expected that their temporal profile will be significantly distorted by the chromatic dispersion of the fibers constituting the NOLM loop. Such profile distortions can significantly alter the TF compared to the analytical form (without dispersion effects) given by the relation (\ref{Pout}). Therefore, here we take into account the fact that in the NOLM loop, the electric fields of the clockwise and counterclockwise waves ($E_{cw}$ and $E_{ccw}$) undergo the combined effects of attenuation, dispersion and Kerr non-linearity, before combining coherently at the exit of the coupler.
However, we assume that the two waves do not interact significantly (by cross-phase modulation) in the NOLM loop, due to their very short duration. In this context, the dynamical behavior of each of the two counter-propagating waves within the NOLM loop can be satisfactorily described by the equation (\ref{eqn:NLSE}). }
This approximation is quite relevant and can be validated by a simple comparison between the experimental results and those of the numerical simulations.
To carry out this comparison, we will use in our simulations the same parameters as those of the incident signal measured experimentally (in Fig. \ref{fig:setup} \hyperref[fig:setup]{(c)} or \hyperref[fig:setup]{(d)}).

\begin{figure}[ht]
\centering\includegraphics[width=.75\linewidth]{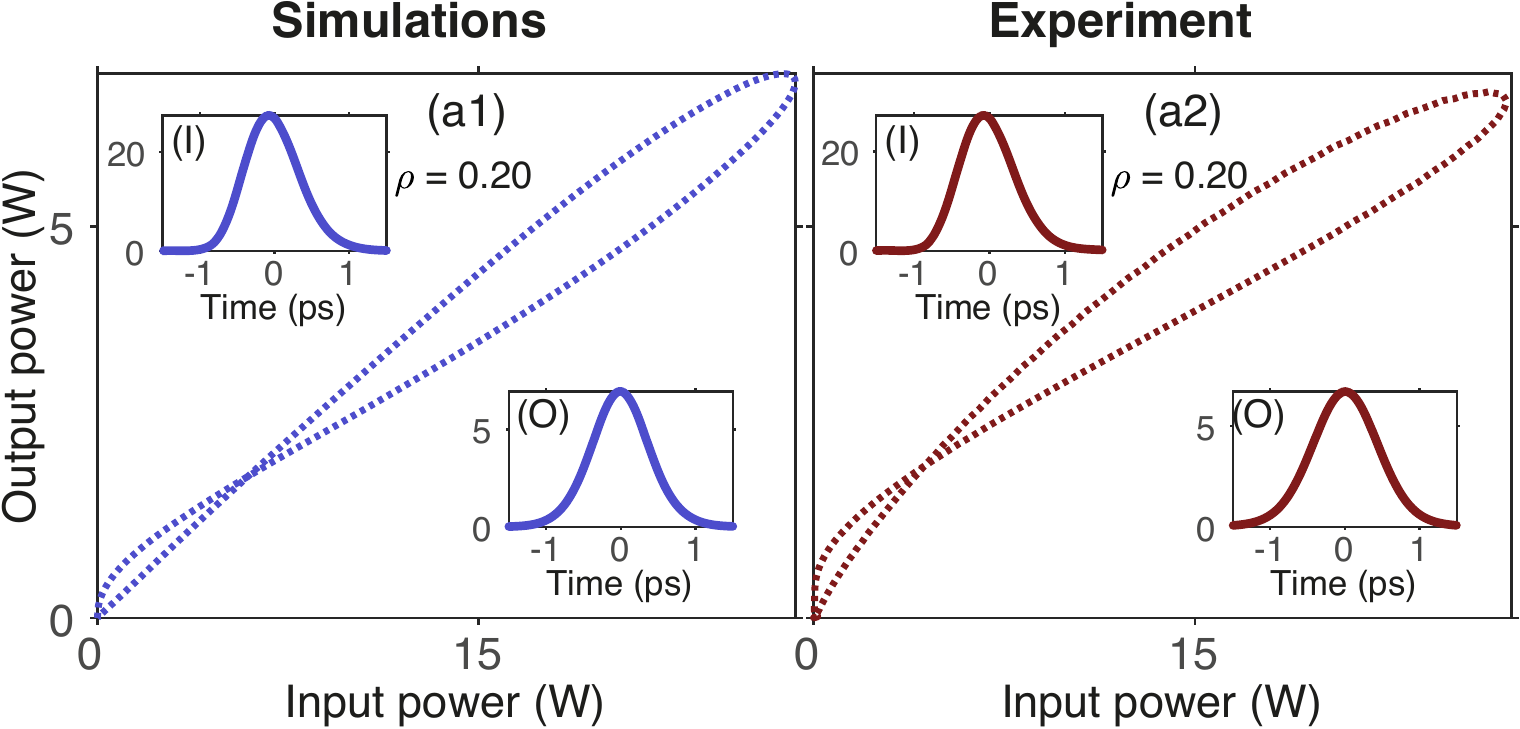}
\caption{Simulated and experimental transfer functions obtained for a coupler ratio of 0.20, and a 2-meter SMF-28 as the NOLM's loop. The inserts labeled "(I)"  and "(O)"  show the temporal profiles of the input and output pulses, respectively.}
\label{fig:2m_SMF}
\end{figure}

To verify the validity of the methodology described above, we first applied it to a simple case, where the loop fiber of the NOLM is a standard 2-meter single-mode fiber (SMF-28).
With a peak power of \SI{27}{\watt}, the nonlinear length in this SMF is equal to \SI{28.5}{\meter}. Because we only use \SI{2}{\meter} of SMF-28, nonlinear effects are thus too weak to change significantly the intensity profile of the signal. 
In this situation, the output signal of the NOLM should be affected only by a reduction in amplitude due to the linear losses and the loss due to the light reflected by the NOLM. Therefore, the TF of the NOLM should be close to a straight line. This is indeed what we can see in the figure \ref{fig:2m_SMF}, both in the results of the numerical simulation and the experiment. The TF exhibits two branches, which result from the asymmetry induced by the TOD between the front and rear edges of the pulse.

Figure \ref{fig:2m_SMF} shows a fairly good agreement between our simulations and the experimental results, thus validating the relevance of our methodology for determining the TF of NOLM.

\section{Results and discussion}

The TF of the NOLM depends on a large number of parameters, including the parameters of the system components and the characteristics of the incident light beam. It is very useful to identify the parameters that have the most impact on the TF and configure the NOLM appropriately according to the application for which it is intended.
One of these parameters is the coupler ratio. Indeed, it is important to bear in mind that any incident beam passing through the NOLM gives rise to two beams, namely, the useful beam, which emerges from the NOLM through the output port, and a reflected beam (which emerges through the input port). This reflected beam constitutes a loss of energy,  which must be controlled with great care in laser cavity mode-locking applications.
The intensity of the reflected beam varies very significantly with the value of the ratio $\rho$, between two limit cases. For $\rho=0$, there is no reflected beam at all, while for $\rho=0.5$ the incident beam is fully reflected. Between these two extreme cases, the intensity of the useful beam exiting the NOLM is all the smaller as the ratio $\rho$ is close to 0.5.
In a NOLM-driven laser cavity, a very low intensity of useful beam would require a huge intra-cavity gain (in mode-locking regime) which can far exceed the gain available in commercial optical amplifiers. In addition, under this operational condition, the curvature of the TF at the origin would curve downwards, with a very low slope. Since the inverse of this slope corresponds to the low-signal gain necessary to compensate for the power loss caused by the NOLM, a very low slope is therefore very likely to interfere with the automatic start of the laser.
In contrast, a NOLM providing a TF with a very low slope at the origin acts as a noise killer, which is suitable for applications related to signal reshaping. 

From now on, we choose the LD-NLF presented in section \ref{sec:NOLM}, to be the loop fiber of the NOLM. However, it is worth noting that the very low dispersion of this fiber can nevertheless have significant effects on picosecond pulses.
In the following we present the TFs measured for different values of the ratio $\rho$, under two operational conditions corresponding to input pulses with relatively large and small temporal widths, respectively.
The results show a very high sensitivity of TF to dispersion effects, despite the low dispersive properties of the NOLM loop fiber.

\subsection{Temporally broadened pulses}

Here, the laser source delivers picosecond pulses (Fig. \ref{fig:setup} \hyperref[fig:setup]{(b)}) that are temporally broadened using a DCF, before being injected into the NOLM.
But before entering the DCF, the pulse's peak power is greatly reduced using an attenuator, in order to minimize the impact of non-linear effects within the DCF.

\begin{figure}[ht]
\centering\includegraphics[width=.7\linewidth]{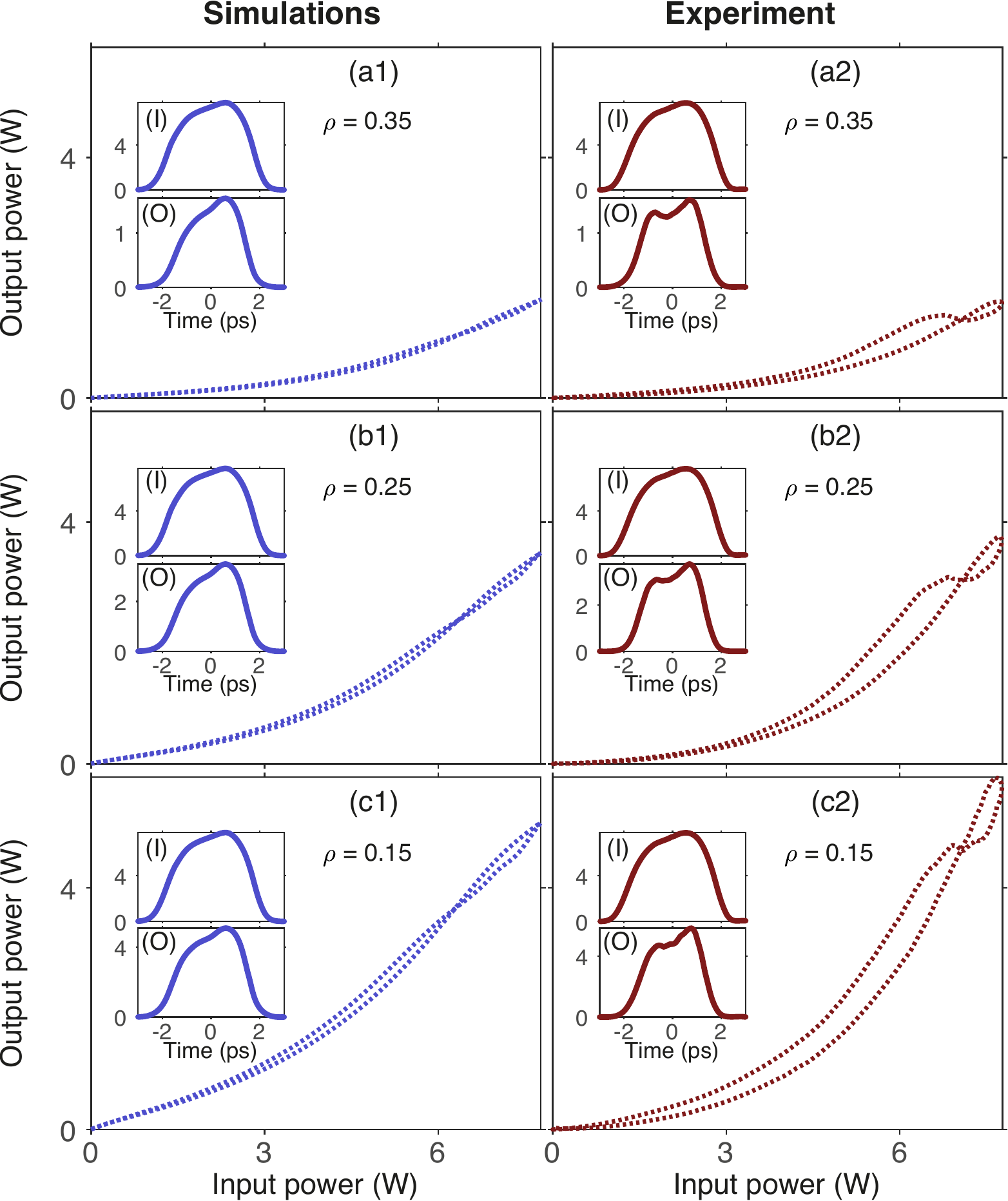}
\caption{Simulated (left) and experimental (right) transfer functions obtained for a coupler ratio of 0.35 (a1-a2), 0.25 (b1-b2), and 0.15 (c1-c2). The inserts labeled "(I)"  and "(O)"  show the temporal profiles of the input and output pulses, respectively.}
\label{fig:Avec_DCF}
\end{figure}

 The incident pulse thus corresponds to the pulse in panel \hyperref[fig:setup]{(c)} of  Fig. \ref{fig:setup}. After this broadening, the pulse peak power is equal to \SI{7.8}{\watt}.  Figure \ref{fig:Avec_DCF} shows several TFs obtained for different value of $\rho$. We can clearly observe in figure \ref{fig:Avec_DCF} that, although the peak power of the input pulse of the NOLM is relatively low, we obtain here a qualitatively different TF from that obtained in Fig. \ref{fig:2m_SMF} corresponding to a NOLM where the loop fiber is the SMF-28. This difference is due to a higher nonlinearity in the LD-NLF, and is particularly evident in the curvature of the TF at the origin, which has a much lower slope in Fig. \ref{fig:Avec_DCF}.
 Moreover, it can clearly be observed in Fig. \ref{fig:Avec_DCF} that
 the slope of the TF at the origin decreases as $\rho$ increases.
We can also observe that at the same time, the output light of the NOLM  decreases (due to the increase in the intensity of reflected light).
Moreover, here too, it can be noted that the TF contains two branches resulting from the asymmetry induced by the TOD of the LD-NLF, but the two branches are relatively close due to a lower impact of the chromatic dispersion on the input pulses considered in Fig. \ref{fig:Avec_DCF}.

\subsection{Case of ultra-short pulses}

\begin{figure}[h]
\centering\includegraphics[width=.75\linewidth]{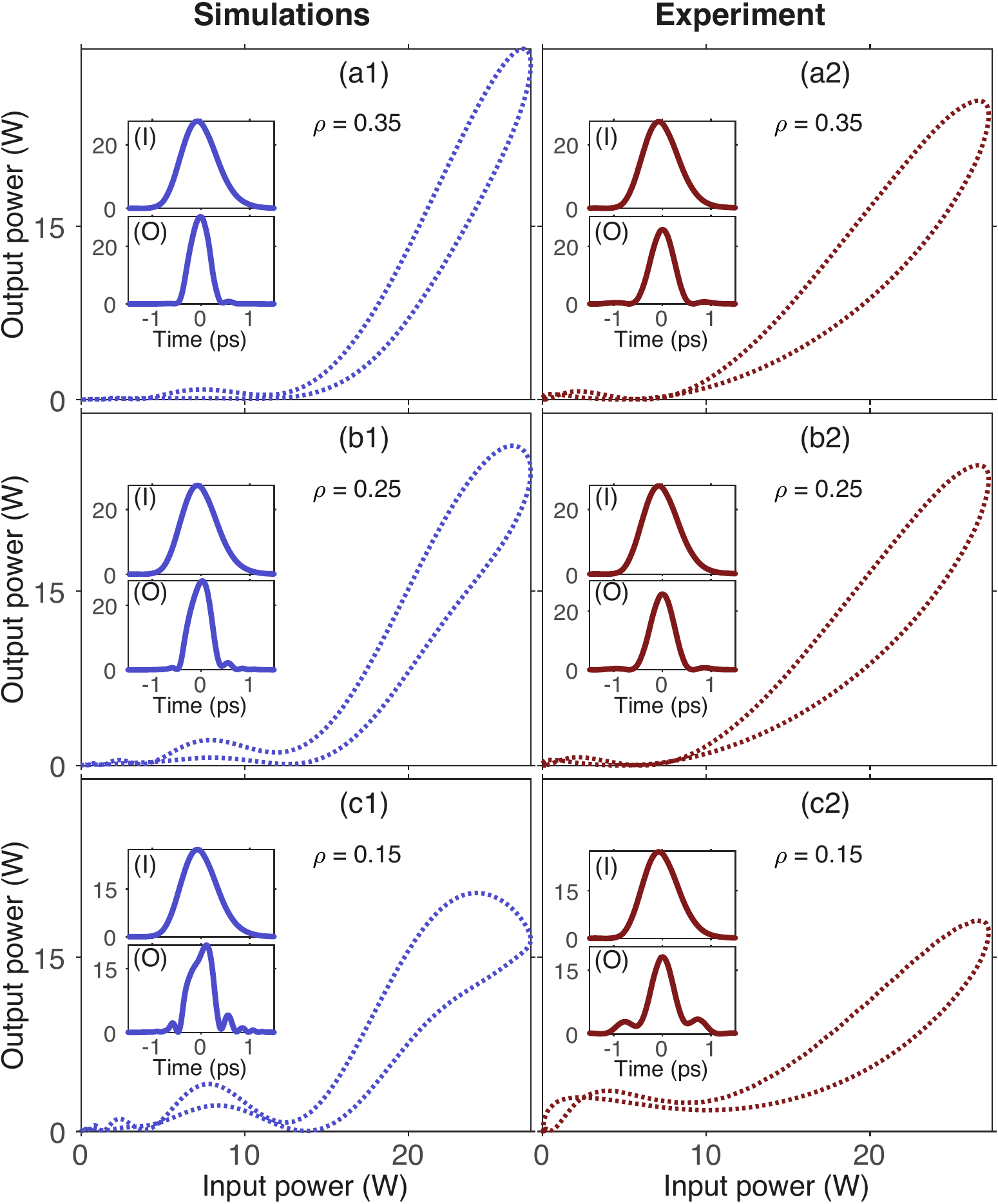}
\caption{Simulated and experimental transfer functions obtained for a coupler ratio of 0.35 (a1-b1), 0.25 (a2-b2), and 0.15 (a3-b3). The inserts labeled "(I)"  and "(O)"  show the temporal profiles of the input and output pulses, respectively.}
\label{fig:Sans_DCF}
\end{figure}

Here, the DCF fiber is removed from our experimental setup (Fig. \Ref{fig:setup}),
and thereby, the incident pulse now has a FWHM of \SI{0.85}{\pico\second},
as can be seen in panel \hyperref[fig:setup]{(d)} of  Fig. \ref{fig:setup}.
The input peak power is increased to \SI{27.2}{\watt}. 
Thus, this input pulse, with a reduced temporal width and an increased peak power compared to the case of Fig. \ref{fig:Avec_DCF}, can be expected to be more prone to the dispersion and non-linearity effects in the LD-NLF.
This is precisely what we can see in all the panels of Fig. \ref{fig:Sans_DCF},
in which the TF now has two distinct branches induced by the TOD. 
It can also be noted in Fig. \ref{fig:Sans_DCF} that the value of $\rho$ has a great impact on the curvature of the TF at the origin. We can see that the slope of the TF at the origin rises as the value of $\rho$ decreases.
On the other hand, a saturation effect is observed around \SI{5}{\watt} for $\rho=0.15$. This saturation effect is weaker when $\rho$ is larger.

\section{Conclusion}

In the present study, we have shown that the TF of a NOLM can be measured experimentally using a FROG system, from temporal measurements of the intensity profiles of the input and output signals of the NOLM.
This type of measurement highlights details to identify the most sensitive parameters of the NOLM, as well as the effects that have the greatest impact on the TF.
In particular, the TOD of the loop fiber of the NOLM has a dramatic and easily identifiable impact, as it produces an asymmetry in the temporal profile of the pulse which creates two branches in the TF.

The experimental results are in fairly good agreement with our numerical simulations, despite the constraints that limit the comparison.
Indeed, our modeling and numerical simulations do not take into account the birefringence effects that are necessarily present in the components we use, which are not polarization maintaining components.
Finally, our method of measuring TFs can be used to verify that the NOLM is optimally configured for applications that require a
 specific TF, such as, for example, to perform mode locking in laser cavities without interfering with the automatic laser start function.

\section{Funding} Agence Nationale de la Recherche (ANR-15- IDEX-
0003, ANR-17-EURE-0002); iXCore Research Foundation; Conseil
régional de Bourgogne-Franche-Comté; FEDER.

\section{Data Availability Statement} Data underlying the results presented in this paper are not publicly available at this time but may be obtained from the authors upon reasonable request.

\section{Disclosure} The authors declare no conflicts of interest.

\bibliographystyle{unsrt} 
\bibliography{Bibliographie_FT_NOLM}

\end{document}